\documentclass[aps,preprintnumbers, nofootinbib, onecolumn, notitlepage]{revtex4-1}
\usepackage{bm}
\usepackage{latexsym}
\usepackage{amsmath,amsfonts,amssymb}
\usepackage{graphicx,epsfig}
\usepackage{psfrag}
\usepackage{amsthm}
\usepackage{color}
\usepackage{amsmath}
\usepackage{braket}
\usepackage{indent first}
\usepackage[normalem]{ulem}
\usepackage{fancyhdr}
\useunder{\uline}{\ul}{}
 \usepackage{environ}
\usepackage{mathtools}
\usepackage{float}
\usepackage{hyperref}
\def\o{\omega}
\def\O{\Omega}

\def\a{\alpha}
\def\b{\beta}

\def\d{\delta}
\def\D{\Delta}

\def\t{\tau}

\def\p{\partial}
\def\f{\frac}

\newcommand{\be}{\begin{equation}}
\newcommand{\ee}{\end{equation}}
\newcommand{\bes}{\begin{equation*}}
\newcommand{\ees}{\end{equation*}}

\newcommand{\beq}{\begin{eqnarray}}
\newcommand{\eeq}{\end{eqnarray}}
\newcommand{\bseq}{\begin{subequations}}
\newcommand{\eseq}{\end{subequations}}

\NewEnviron{eqn}{
\begin{align}
\begin{split}
  \BODY
\end{split}
\end{align}
}

\NewEnviron{eqn*}{
\begin{align*}
\begin{split}
  \BODY
\end{split}
\end{align*}
}

\begin{document}
%
%
%
%
\title{{\bf Discreteness of Space from GUP in Strong Gravitational Fields}}
%
%
%
%
\author{Ashmita Das}
\email{ashmita.phy@gmail.com}
\affiliation{Department of Physics, Indian Institute of Technology Madras, Chennai 600036, India}

\author{Saurya Das}
\email{saurya.das@uleth.ca}
\affiliation{Theoretical Physics Group and Quantum Alberta, Department of Physics and Astronomy, University of Lethbridge, 4401 University Drive, Lethbridge, Alberta T1K 3M4, Canada}

\author{Elias C. Vagenas}
\email{elias.vagenas@ku.edu.kw}
\affiliation{Theoretical Physics Group, Department of Physics, Kuwait University, P.O. Box 5969, Safat 13060, Kuwait}
%
%
%
%
\begin{abstract}
%
%
%
%
\par\noindent
A large class of quantum theories of gravity show that the Heisenberg's uncertainty principle is modified to the ``Generalised Uncertainty Principle" (GUP) near the Planckian scale. It has also been shown that the GUP induces perturbative corrections to all quantum mechanical Hamiltonians, even at low energies, and thereby introduces Planck scale
corrections to the Schr\"odinger equation and to the relativistic quantum mechanical equations. Some of these corrections give rise to potentially measurable effects in the low-energy laboratory.
Another prediction of these corrections is that 
a measured length must be quantized,
as seen by studying the solutions of the GUP modified
Schr\"odinger, Klein-Gordon, and Dirac equations in 
a one, two, and three dimensional box.
This result was subsequently extended to spacetimes with weak gravitational fields. 
In this work, we further extend this length quantization to spacetimes with strong gravitational fields and show that this result continues to hold,  thereby showing that it is robust. 
\end{abstract}
\maketitle
%
%
%
%
%
%
\section{Introduction}
%
%
%
%
%
%
\par\noindent
It is well known that the standard quantum field theory (QFT) is successful up to a certain energy scale $\Lambda \ll M_{Pl}c^2$, where $M_{Pl}$ is the Planck mass and $M_{Pl}c^2 \sim 10^{19}~GeV$ is the Planck energy. 
As a result, QFT generally excludes the energy scale associated with quantum gravitational interactions. 
A related issue is that in quantum gravity (QG), one expects quantum fluctuations of the background spacetime, whereas standard QFT assumes a fixed background spacetime on which quantum fluctuation of matter and gauge fields are studied. 
As a result, a complete and consistent theory of QG is yet to be formulated, although there are promising candidate theories such as String Theory,  Loop Quantum Gravity, Causal Dynamical Triangulations, Doubly Special Relativity (DSR) etc. 
A consistent theory of QG may be a gateway to the unification of all fundamental forces of nature. 
One shortcoming of all such theories is the complete absence of direct or
indirect experimental evidence in support of or contradicting them. 
This being clearly undesirable, it is important to explore potential signatures of these theories in current or future experiments. This has been the subject of study of Quantum Gravity Phenomenology \cite{AmelinoCamelia:2008qg, Hossenfelder:2012jw}. 

In this respect, one has often exploited a 
robust prediction of candidate theories of QG,
namely a minimum uncertainty in position measurement to $\mathcal{O}$($l_{Pl}$) with  $l_{Pl}\sim 10^{-35}~m$ to be the Planck length, the corresponding  existence of a minimum measurable length \cite{Amati:1988tn,Garay:1994en,Kempf:1994su}, and the modification of the Heisenberg uncertainty principle (HUP) to the so-called Generalised Uncertainty Principle (GUP) \cite{Maggiore:1993rv,Maggiore:1993zu,Maggiore:1993kv,Scardigli:1999jh}.

The GUP implies a modification of the standard Heisenberg commutator, i.e., $[x_i,p_j]=i\hbar\delta_{ij}$,  by 
terms induced from QG.
Although these terms imply significant contributions near the Planck length or energy scale, they give rise to potentially measurable effects at much lower energy scales. 
A generalization of the GUP was proposed in Refs. \cite{Das:2008kaa, Das:2010sj,Das:2010sj,Das:2009hs,Ali:2009zq,
Ali:2010yn,Das:2010sj} of the form  
\bseq
\begin{align}
[x_i,\,p_j]=i\hbar\bigg[\d_{ij}-&\a\bigg(p\,\d_{ij}+\f{p_ip_j}{p}\bigg)+\a^2\bigg(p^2\,\d_{ij}+3\,p_ip_j\bigg)\bigg]\label{nrgup_1}\\[10pt]
&[p_i,\,p_j]=[x_i,\,x_j]=0 ~.\label{nrgup_1_a}
\end{align}
\eseq
This  modification of the standard Heisenberg commutator corresponds to the following modified position-momentum uncertainty relation 
\beq
\D x\,\D p &\geqslant&\,\f{\hbar}{2}\bigg[1-2\a\braket{p}+4\a^2\braket{p^2}\bigg]\nonumber\\
&\geqslant&\,\f{\hbar}{2}\bigg[1+\bigg(\f{\a}{\sqrt{\braket{p^2}}}+4\a^2\bigg)\,\D p^2+4\a^2\braket{p}^2-2\a\sqrt{\braket{p^2}}\bigg]
\label{nrgup_2}
\eeq
where $i$ and $j$ take values $1,2,3$, the magnitude of the squared momentum is $p^2=\sum_{j=1}^{3}\,p_{j}p_{j}$, and  $\a=\a_0/M_{Pl}c=\a_0l_{Pl}/\hbar$ with  $\a_0$ to be a dimensionless constant, sometimes assumed to be 
${\cal O}(1)$\footnote{Recently, in Ref. \cite{Scardigli:2016pjs}, using a GUP with only a quadratic term in momentum, a numerical value of the dimensionless GUP parameter $\beta_0$ (with $\beta_0 \sim\alpha^{2}_{0}$)
was obtained, namely $82\pi/5$. Furthermore, in Ref. \cite{Vagenas:2018zoz} a similar analysis 
with linear $+$ quadratic GUP yielded $\alpha_{0}$ to be proportional to powers of the dimensionless ratio ($M_{Pl}$/M).}.
since GUP as formulated allows for a general $\alpha_0$ and one does not yet have a direct measurement of quantum gravity parameters, we leave it arbitrary. This in turn implies the existence of an intermediate length scale and imposes meaningful bounds on the quantum gravity parameters, which can in principle be measurable in future experiments \cite{Das:2008kaa, Das:2010sj, Das:2009hs,Ali:2009zq}.

\par\noindent
The above form of GUP is compatible with the modification of uncertainty principle, as proposed in String Theory, DSR, and black hole (BH) physics. For alternate forms of GUP, we refer our readers to Refs. \cite{Kempf:1994su,Kempf:1996fz,Brau:1999uv,Hossenfelder:2003jz,Das:2008kaa, Adler:1999bu, Scardigli:2003kr, Jizba:2009qf} and for phenomenological implications to Refs. \cite{Ali:2010yn,Ali:2011fa,Basilakos:2010vs,Scardigli:2014qka}.
%
%
%
%
Note that having been derived as a consequence of DSR theories, although the form of GUP as given in Eq. (\ref{nrgup_2}) is not  manifestly Lorentz invariant, it is covariant under the non-linear Lorentz transformations in DSR theories \cite{Magueijo:2001cr, AmelinoCamelia:2000mn, AmelinoCamelia:2000ge, Magueijo:2002am, Cortes:2004qn}.
The aforementioned forms of the modified commutation relation, namely Eq. (\ref{nrgup_1}), and of the GUP, namely Eq. (\ref{nrgup_2}), imply a minimum measurable length and a maximum measurable momentum of the form
\bseq
\begin{align}
\D x\,\geqslant\,(\D x)_{{\rm min}}\, \approx\, \a_0\,l_{Pl}\\
\nonumber\\
\D p\,\leqslant\,(\D p)_{{\rm max}}\,\approx \,\f{M_{Pl}\,c}{\a_0}~.
\end{align}
\eseq
For ease of calculations, one defines the physical position and momentum operators (which are no longer canonically conjugate) 
in terms of auxiliary and ``canonical'' variables $x_{0i}$ and $p_{0i}$, such that 
\begin{eqnarray}
&& x_i\,=\,x_{0i},\,\,\,\,\,\,\,\, p_i=p_{0i}\left(1-\a p_0+2\a^2 p_{0}^{2}\right) \label{mod_op_1} \\
&& [x_{0i}, p_{0j}] \,=\,i\hbar\, \delta_{ij}
\end{eqnarray}
\par\noindent
with  $p_{0i}=-i\hbar\f{\p}{\p x_{0i}}$ to be interpreted as the components of the low-energy momentum $p_{0}$, where the QG effects are negligible, and $p_{0}^{2} \equiv \sum_{i=1}^{3}p_{0i}p_{0i}$. On the other hand, $p_i$ may be regarded as the momentum at ``high energies". Using the above equations, one can indeed show that Eq. (\ref{nrgup_1}) is satisfied.
\par\noindent
In the current paper, we re-examine an interesting aspect of 
GUP, which is the prediction that not only is there a minimum length, but 
that all measurable lengths are quantized and discrete. Note that the former does not automatically imply the latter. A similar set of results follows for
area and volume measurements as well. It has been shown earlier that 
this discreteness holds not only for flat background spacetimes, but also
in the presence of weak gravity. 
%
%
%
%
Specifically, in Ref. \cite{Ali:2009zq}, the authors have shown that the one-dimensional space confining an elementary particle must be discrete, implying all measurable lengths are quantized in units of a fundamental length scale, which can be the Planck length. Subsequently,  in Ref.  \cite{Das:2010zf}, this was extended for a relativistic particle confined in a rectangular as well as in a spherical box in one, two, and three dimensional space by solving the GUP-modified Klein Gordon and Dirac equations.  It was shown that the length, area, and volume of the box are quantized in units of a fundamental length scale. Extending the results of flat spacetime to weakly curved spacetime, the authors of Ref. \cite{Deb:2016psq} have shown that the quantization of lengths, areas, and volumes continue to hold in the weak gravitational regime. These results support the fundamentally discrete nature of space and a breakdown of the spacetime continuum picture near the Planck scale. 
Here we will explore the extent to which the length quantization holds for strong gravity, i.e., in the background of spacetimes of high curvature. As a concrete example of a spacetime with strong gravity, we first consider the Schwarzschild BH metric in its high curvature region. Without exploiting any other BH property such as its horizon or singularity, we show that our result for the quantization of the length of the box continues to hold. Then, we obtain similar results for the case of Reissner-Nordstr\"om (RN) BH and  for the case of the cosmological Friedmann-Lema$\hat{{\rm i}}$tre-Robertson-Walker (FLRW) spacetime. 
In the above background spacetimes, we apply the aforementioned GUP, which is covariant under the non-linear Lorentz transformation. 
%
Following our earlier works, our strategy would be to
solve the scalar field equation in the background of
the aforementioned spacetimes, impose ``box boundary conditions'' such that the field is confined to a finite region of space and show that such a 
region can only be of quantized or discrete size.
In other words, the length measurements, which require the confinement of one or multiple particles (described by the scalar field) within a boundary, signifying the end points of the length to be measured, imply in turn that the measurable lengths are discrete as opposed to continuous.  Deriving this in the context of strong gravity completes this program.  Although discreteness of lengths, areas, and volumes have been derived from other approaches in the past, our approach here for the length quantization is relatively simpler and does rely neither on any particular approach of quantum gravity, nor does it require additional assumptions \cite{Thiemann:1996at,Rovelli:1994ge,Nicolai:2005mc}.
\par\noindent
The remainder of the paper is organized as follows. In sections \ref{metric_singular}, \ref{gup_rn}, and \ref{gup_flrw} we present the analysis for the quantum character of space in the strong gravity regions, as produced by the Schwarzschild BH, RN BH, and FLRW metrics, respectively.
Finally, in section {\ref{discussions}}, we discuss our findings and conclude. Appendices are provided for the paper to be self-sufficient. Throughout the paper, we consider the metric signature to be $(-,+,......+)$.
%
%
%
%
%
 \section{GUP in  Schwarzschild BH spacetime}
 \label{metric_singular}
%
%
%
%
\par\noindent
In this section we study the scalar field equation in the strong gravitational field of the Schwarzschild BH spacetime. We start working in $D$ spacetime dimensions  with coordinates  ($t,r,\chi_1,\chi_2,...\chi_{D-2}$) and, thus, the Schwarzschild BH metric reads \cite{Myers:1986un}
\begin{eqnarray}
&& ds^2=-f(r)c^{2}dt^2+\frac{1}{f(r)}dr^2+r^2d\Omega^{2}_{D-2}
\label{genmetric1}
\end{eqnarray}
with  $f(r)=\bigg(1-\frac{2\,GM}{c^2\,r^{D-3}}\bigg)$ and 
$d\O_{D-2}^2=d\chi_{1}^{2}+{\rm sin}^{2}\chi_{1}\,d\chi_{2}^{2}+{\rm sin}^{2}\chi_{1}\, {\rm sin}^{2}\chi_{2}\, d\chi_{3}^{2}+{\rm sin}^{2}\chi_{1}.....{\rm sin}^{2}\chi_{D-3}\, d\chi^{2}_{D-2}$, where $\Omega_{D-2}$ is the 
area of a unit sphere $S^{D-2}$. The radius of the event horizon is given as $r_H=\big(\f{2\,GM}{c^2}\big)^{1/(D-3)}$  with $M$ to be the BH mass. 
It should be stressed that the above BH spacetime consists of three distinct regions:  (i) the asymptotically flat region, i.e., $r\to \infty$, (ii) the region near the event horizon of the BH, i.e., $r\sim r_H$, and (iii) the region near the BH curvature singularity, i.e., $r\to 0$. Region (iii) is the strong gravity region in which the metric given by Eq. (\ref{genmetric1}) reduces to
\begin{equation}
ds^2=\frac{2\,GM}{c^2\,r^{D-3}}c^{2}dt^2-\frac{c^2\,r^{D-3}}{2\,GM}dr^2+r^2 d\Omega^{2}_{D-2}~.
\label{metsingularity}
\end{equation}
At this point it should be stressed that we will keep the initial formalism general for any $D$ since this is straightforward and it may be useful for future generalizations of our work. We will then specialize to $D=4$.
\\
It is well known that the scalar field or Klein-Gordon (KG) equation in any $D$-dimensional curved spacetime is written as
\begin{eqn}
(\Box +\f{m^2c^2}{\hbar^2})\Phi=0\label{scalar1}~.
\end{eqn}
Assuming $\Phi(t,r,\chi_1,\chi_2,\ldots,\chi_{D-2})=\t(t)\psi(r)\,Y_{l,m,n,\ldots}(\chi_1,\chi_2,\chi_3,\ldots,\chi_{D-2})$
and separating variables, 
Eq. (\ref{scalar1}) now reads (see Appendix A)
\beq
\underbrace{-\f{\hbar^2}{r^{D-2}}\f{d}{dr} \bigg[r^{D-2}\f{d\psi(r)}{dr} \bigg]}_{p_{0}^{2}}\,\,f(r)-\hbar^2\,f'(r)\f{d\psi}{dr}+\bigg[\frac{\hbar^2\,l(l+1)}{r^2}-\, \f{E^2}{c^2\,f(r)}+m^2c^2\bigg]\psi(r)=0
\label{scalar10}
\eeq
where $l=0,\,1,\,2\,\ldots$ is the orbital quantum number and $E$ the energy of the scalar field. 
It is easily seen that Eq. (\ref{scalar10}) portrays the radial equation of motion for a scalar field in the Schwarzschild BH background. At this point we need to modify the squared low-energy 3-momentum, i.e., $p_{0}^{2}$, due to the GUP and for this reason we  utilize Eq. (\ref{mod_op_1})\footnote{The angular part can be GUP modified as well \cite{Bosso:2016frs}, although since it does not contain any black hole parameters, it is insensitive to them. Therefore, the result, i.e., discreteness of space in those directions, would not differ from those found in flat spacetimes as well as in weak gravitational backgrounds \cite{Ali:2009zq, Ali:2011fa, Das:2010zf, Deb:2016psq}.}.  Therefore, the squared low-energy 3-momentum, i.e., $p_{0}^{2}$, will be transformed to the high-energy 3-momentum, i.e., $p_{i}^{2}$, as follows \cite{Ali:2009zq, Das:2010zf}
\be
p_{0}^{2}\to\,p_{i}^{2}=-\hbar^2\, \nabla^2\,\psi(r)-2\,i\a\hbar^3\, \nabla^2(\nabla \psi(r))+5\a^2\hbar^4 \, \nabla^2(\nabla^2 \psi(r))\label{gup_mom}
\ee
where the Laplace operator is written in  the Cartesian coordinates. Hence, we transform the operator to the spherical polar coordinates and so we  use $p_{i}^{2}$ in Eq. (\ref{scalar10}). Thus, we obtain the GUP modified scalar field equation 
\beq
&-\f{\hbar^2}{r^{D-2}}\f{d}{dr} \bigg[r^{D-2}\f{d\psi(r)}{dr} \bigg]-\f{2\,i\a\hbar^3}{r^{D-2}}\,\bigg(\f{d}{dr}\bigg[r^{D-2}\f{d}{dr}\bigg]\f{d\psi}{dr}\bigg)+\f{5\a^2\hbar^4}{r^{D-2}}\f{d}{dr}\bigg[r^{D-2}\f{d}{dr}\bigg]\bigg[\f{1}{r^{D-2}}\f{d}{dr} \bigg\{r^{D-2}\f{d\psi(r)}{dr}\bigg\} \bigg]\nonumber\\
&-\f{\hbar^2\,f'(r)}{f(r)}\f{d\psi}{dr}+\bigg[\frac{\hbar^2\,l(l+1)}{f(r)\,r^2}-\, \f{E^2}{c^2\,f^2(r)}+\f{m^2c^2}{f(r)}\,\bigg]\psi(r)=0~.
\label{scalar12}
\eeq
The above equation can be viewed as the master equation for the massive KG field in the background of a stationary and spherically symmetric spacetime. 
\par\noindent
Now we focus on a $4$-dimensional spacetime, namely $D=4$, for which Eq. (\ref{scalar12}) reduces to
\begin{eqnarray}
-\f{2\hbar^2}{r}\psi'(r)-\hbar^2\,\psi''(r)-&\f{4i\a\hbar^3}{r}\,\psi''(r)-2i\a\hbar^3\,\psi'''(r)+\f{20\a^2\hbar^4}{r}\,\psi'''(r)+5\a^2\hbar^4\,\psi''''(r)\nonumber\\
&-\f{\hbar^2\,f'(r)}{f(r)}\,\psi'(r)+\bigg[\frac{\hbar^2\,l(l+1)}{f(r)\,r^2}-\, \f{E^2}{c^2f^2(r)}+\f{m^2c^2}{f(r)}\,\bigg]\psi(r)=0~.
\label{scalar_4d_1}
\end{eqnarray}
\par\noindent
Utilizing the following approximations in the $r\rightarrow 0$ limit
\beq
\lim_{r\to 0}f(r)\,\approx\,\bigg(-\f{2GM}{rc^2}\bigg)\,\,\,\,\,\,\,\,\,\,\,\,\,\,\,\,\,\,\,\,\,
\lim_{r\to 0}f'(r)\approx \,\bigg(\f{2GM}{c^2r^2}\bigg)\,\,\,\,\,\,\,\,\,\,\,\,\,\,\,\,\,\,\,\,\,\,\,\,\,\,
\lim_{r\to 0}\f{f'(r)}{f(r)}\approx \,-\f{1}{r}
\label{f(r=0)}
\eeq
and omitting ${\cal O}(\alpha^2)$ terms, Eq. (\ref{scalar_4d_1}) after taking the limit $r\to 0$, reduces to
\be
-\hbar^2\,\psi'(r)-4i\a\hbar^3\,\psi''(r)+\bigg[\f{c^2\hbar^2\,l(l+1)}{-2GM}\bigg]\,\psi(r)=0~.
\label{scalar_4d_sing_2}
\ee
The above equation has a solution of the form \footnote{For $\a \to 0$, Eq. (\ref{scalar_4d_sing_2}) becomes a first order differential equation and its solution contains one unknown constant.  It can be shown by imposing $\a \to 0$ limit in Eq. (\ref{sol_4d_1}), which is the solution for second order differential Eq. (\ref{scalar_4d_sing_2}), that only the term associated with constant $C_1$ matches exactly with the solution of the first order differential equation.}
\beq
\psi(r)=C_1\, {\rm exp}\bigg[\f{ir}{8\a\hbar}\bigg\{1-\bigg(1-\f{8ic^2\,\a\hbar\,l(1+l)}{GM}\bigg)^{1/2}\bigg\}\bigg]
+C_2\, {\rm exp}\bigg[\f{ir}{8\a\hbar}\bigg\{1+\bigg(1-\f{8ic^2\,\a\hbar\,l(1+l)}{GM}\bigg)^{1/2}\bigg\}\bigg]~.
\label{sol_4d_1}
\eeq
Now we impose the boundary conditions on solution given by Eq. (\ref{sol_4d_1}). To determine the boundary conditions near  the $r \to 0$ region of the Schwarzschild BH, we adopt a similar setup as for a particle confined in a box. 
Due to the spherical symmetry of the  Schwarzschild BH, we imagine a black sphere of radius $L$ inside the event horizon of BH, namely $L<\,r_H$. Thus, the boundary conditions turned out to be 
\be
\psi(r=0)\,=\,0\,\,\,\,\, \mbox{and}\,\,\,\,\, \psi(r=L)=0~.\label{bdcs}
\ee
At this point it should be stressed that at $r=0$ a physical (curvature) singularity occurs. However, normalizable wavefunctions should remain finite there and therefore the above boundary condition at $r=0$ is justified. 
Furthermore, as we show in Appendix B, if one sets  $\psi(r=\epsilon)=0$ with $\epsilon\approx 0$ and then take the limit $\epsilon\rightarrow 0$, the obtained quantization condition is identical with the one as obtained here by setting $\psi(r=0)=0$.


%
%
%
%
\par\noindent
The boundary condition at $r=0$ on the wave function Eq. (\ref{sol_4d_1}) yields
\be
C_1=\,-\,C_2 \label{bdc_1}
\ee
\par\noindent
which now reads
%
%
%
%
\beq
\psi(r)=\,\tilde{C}\,{\rm exp}\bigg(\f{i\rho}{\a}-\f{i\pi}{2}\bigg)\, \sin\bigg[\f{\rho}{\a}\,(1-i\a\b)^{1/2}\bigg]
\label{sol_4d_5}
\eeq
\par\noindent
where  $\tilde{C}=2C_1 \equiv |\tilde{C}|\,{\rm exp}(-i\,\theta_{\tilde{c}})$, $\b=\f{16\hbar\,l(l+1)}{r_H}$, and $\rho=\f{r}{8\hbar}$. 
Next, we impose the boundary condition at $r=L$ on the wave function given by Eq. (\ref{sol_4d_5}) and this yields 
\be
\psi(r=L)=|\tilde{C}|\,{\rm exp}\bigg(\f{iL}{8\a \hbar}-\f{i\pi}{2}-i\theta_{\tilde{c}} \bigg)\, \sin\bigg[\f{L}{8\a \hbar}\,(1-i\a\b)^{1/2}\bigg]
=0~.
\label{sol_4d_6}
\ee
\par\noindent
This leads to two possibilities:
%
%
\beq
{\rm Case\,\,1\, :} \,\,\,\,\,\,\,\,\,\,\,\,\,&{\rm exp}\bigg(\f{i\,L}{8\a \hbar}-\f{i\pi}{2}-i\theta_{\tilde{c}}\bigg)=0\nonumber\\
{\rm Case\,\,2\, :}\,\,\,\,\,\,\,\,\,\,\,\,\,
&\sin\bigg[\f{L}{8\a\hbar}\,(1-i\a\b)^{1/2}\bigg]=0~.
\eeq
\par\noindent
{Case 1 does not give rise to any consistent condition while,
%
%
%
%
%
%
%
from Case 2,  we obtain}
\beq
&&\bigg[\f{L}{8\a\hbar}-\,\f{L}{8\hbar}\f{i\b}{2}+\,\f{L}{8\hbar}\f{\a\b^2}{8}+O(\a^2)\bigg]\,=\,n\pi\nonumber\\
&&\implies\,\f{L}{8\a\hbar}\,\bigg[1+\,\f{\a^2\b^2}{8}\bigg]\,=\,n\pi \nonumber\\
&&\implies\, \f{L}{8\a\hbar}\,\approx \,n\pi\,\bigg[1-32\,\f{\a_{0}^{2}\,l_{pl}^{2}\,l^2(l+1)^2}{r_{H}^{2}}\bigg]\nonumber\\
&&\implies\, \f{L}{8\a\hbar}\,\approx \,n\pi~.
\label{bdc_pos_2}
\eeq
%
%
{
Therefore, omitting $O(\a^2)$ terms, we end up with a length quantization condition of the form
\be
L= 8n\pi \alpha_{0}l_{pl}
\label{quantz_small_r}
\ee
with $n\,\in\,\mathbb{N}$.}
\par\noindent
A number of comments are in order. First, the length quantization condition is independent of the curvature of spacetime 
and the mass of the test field. This makes the length quantization condition robust and trustworthy. 
Second, starting from Eq. (\ref{scalar_4d_sing_2}), we have been mostly working only with the linear order 
of the GUP parameter, i.e.,  $\a$. For Case 2 here, 
we retained the $O(\a^2)$ term  until we obtain the condition in Eq. (\ref{bdc_pos_2}) to demonstrate that the term includes gravitation/curvature in terms of the Schwarzschild BH radius $r_H$. Third, our result suggests that the space in a strong gravity region of the Schwarzschild BH spacetime is indeed quantized in units of a fundamental length scale and the discrete nature of space continues to hold.\\
%
%
%
%
%
%
\section{GUP in RN BH spacetime}\label{gup_rn}
%
%
%
%
%
%
\par\noindent
In this section we study the scalar field equation in the strong gravitational field of the RN
spacetime.  
The $4$-dimensional RN metric in spherical polar coordinates can be described as
\beq
ds^2=-f(r)c^{2}dt^2+\frac{1}{f(r)}dr^2 +r^2\,d\O^2\label{metric_rn_1}
\eeq
%
%
%
%
%
%
%
with 
\beq
f(r)=\bigg(1-\f{r_s}{r}+\f{r_{Q}^{2}}{r^2}\bigg),\,\,\,\,\,\,\,\,\,\,\,\,\,\,\,\,r_s=\f{2GM}{c^2},\,\,\,\,\,\,\,\,\,\,\,\,\,\,\,\,\,r_{Q}^{2}=\f{Q^2\,G}{4\pi \epsilon_0\,c^4} \label{relation_rn_1}~.
\eeq
%
%
%
%
%
\par\noindent
Using the following approximations in the $r\to 0$ limit for the RN metric, we obtain 
\beq
\lim_{r\to 0}f(r)\,\approx\,-\f{r_s}{r}\bigg(1-\f{r_{Q}^{2}}{r_s\,r}\bigg)\,\approx\,\,\f{r_{Q}^{2}}{r^2},\,\,\,\,\,\,\,\,\,\,\,\,\,\,\,\,\,\,\,\,\,
\lim_{r\to 0}f'(r)\approx \,-\bigg(\f{2r_{Q}^{2}}{r^3}\bigg),\,\,\,\,\,\,\,\,\,\,\,\,\,\,\,\,\,\,\,\,\,\,\,\,\,\,
\lim_{r\to 0}\f{f'(r)}{f(r)}\approx \,-\,\f{2}{r}~.
\label{f(r=0)_rn}
\eeq
For the scalar field in RN background one can indeed use the master equation for a 4-dimensional spacetime as given by Eq. (\ref{scalar_4d_1}). Thus, using Eq. (\ref{f(r=0)_rn}) into the master equation and implementing the approximations for $r\to 0$ limit as adopted in case of Schwarzschild BH, we get\footnote{Since we study the high curvature region of the black hole spacetimes, namely $r\rightarrow 0$, in this regime, the Reissner-Nordstr\"om metric (which goes as $1/r^2$) does not smoothly go over to the Schwarzschild metric (which goes as $1/r$). This is the reason why Eqs.(14) and (26) are different and one cannot smoothly go from one to the other.}
\beq
4i\,\a\hbar^3 \psi''(r)\,-\,20\a^2\,\hbar^4\,\psi'''(r)\,=\,0~.
\label{kg_rn_strong}
\eeq
It is evident that in Eq. (\ref{kg_rn_strong}) if we keep terms up to the linear order in $\a$, no GUP modifications will show up. This leads us to keep terms up to the quadratic order of $\a$ in the GUP modified KG equation. 
Therefore, we write
\beq
i\, \psi''(r)\,-\,5\a\,\hbar\,\psi'''(r)\,=\,0
\label{kg_rn_strong_1}
\eeq
and solving the above equation, we obtain
\beq
\psi(r)=\,(-25\,\a^2\,\hbar^2\,C_1)\,e^{ir/5\,\a\,\hbar}\,+\,C_2\,r\,+\,C_3~.
\label{sol_rn_1}
\eeq
Therefore, in case of the KG equation in the background of the strong gravity region of the RN spacetime, the corrections due to GUP start to emerge from the quadratic order of $\a$. This feature distinguishes the strong gravity region of 
RN from that of Schwarzschild spacetime. 
\par\noindent
Following a similar analysis as in section \ref{metric_singular}, we implement the boundary condition at $r=0$ and obtain the  condition 
\beq
C_3=\, 25\,C_1\a^2\,\hbar^2~.
\label{bdc_rn_1}
\eeq
Then, implementing the boundary condition at $r=L$, we obtain
\beq
C_2\,L+\,25\,C_1\a^2\,\hbar^2\,\bigg[1-\,{\rm cos}\bigg(\f{L}{5\,\a\hbar}\bigg)\bigg]+i\,25\,C_1\a^2\,\hbar^2\,{\rm sin}\bigg(\f{L}{5\,\a\hbar}\bigg)\,=\,0~.
\label{bdc_rn_2}
\eeq
Equating the imaginary parts from both sides of Eq. (\ref{bdc_rn_2}), we obtain a quantization condition similar to Eq. (\ref{bdc_pos_2})
\beq
&&{\rm sin}\bigg(\f{L}{5\,\a\hbar}\bigg)\,=\,0\nonumber\\
&&\implies\, \f{L}{5\,\a\hbar}\,=\,n_{1}\pi
\label{quan_cond_rn_1}
\eeq
where $n_{1}\,\in\,\mathbb{N}$. 
\par\noindent
Equating the real parts from both sides of Eq. (\ref{bdc_rn_2}), we obtain  the following condition 
\beq
{\rm cos}\bigg(\f{L}{5\,\a\hbar}\bigg)\,=\,1+\f{C_2\,L}{25\,C_1\a^2\,\hbar^2}~.
\label{quan_cond_rn_2}
\eeq
The LHS of the above equation contains the even powers of $L$, thus, two conditions emerge of the following form
\beq
{\rm condition}\,\, 1:\,\,&&{\rm cos}\bigg(\f{L}{5\,\a\hbar}\bigg)\,=\,1\implies\, \f{L}{5\,\a\hbar}\,=\,2\,n_{2}\,\pi
\label{quan_cond_rn_3}\\
{\rm condition}\,\, 2:\,\,&&\,\f{C_2\,L}{25\,C_1\a^2\,\hbar^2}\,=\,0
\label{quan_cond_rn_4}
\eeq
where, $n_{2}\,\in\,\mathbb{N}$. From condition 2,  $C_2$ turns out to be zero as $L \neq 0$. 
\par\noindent
{Combining the conditions Eqs. (\ref{quan_cond_rn_1}) and (\ref{quan_cond_rn_3}) which have to be satisfied simultaneously, we end up with a length quantization condition of the form 
\be
L= 10 n\pi \alpha_{0}l_{pl}
\label{quantz_small_r}
\ee
with $n\,\in\,\mathbb{N}$.}
Our result suggests that the space in a strong gravity region of the RN BH spacetime is indeed quantized in units of a fundamental length scale and the discrete nature of space continues to hold.
%
%
\section{GUP in FLRW spacetime}\label{gup_flrw}
%
%
%
%
%
\par\noindent
In this section we study the scalar field equation in the background of the FLRW spacetime.
Studying the KG equation in the background of FLRW spacetime will lead us to the discrete nature of space as well, as we
shall see. 
\par\noindent
We consider the observed spatially flat FLRW metric in spherical polar coordinates 
\beq
ds^2= -c^2dt^2+\,a^{2}(t)\bigg[dr^2+\,r^2\,d\O^2\bigg]
\label{metric_flrw}
\eeq
where $a(t)$ is the scale factor of the Universe. 
As is well known, the above spacetime correctly describes the Universe 
at large scales for all epochs except the very early Planck epoch.
This includes the 
inflationary epoch, radiation (RD) and matter dominated (MD) phases, as
well as the current accelerated expanding phase.
To identify the strong gravity regime for the FLRW spacetime, we first do a comparative study of the various phases starting from the inflationary epoch to the MD epoch and estimate the scalar curvature in each epoch. 
%
%
%
%
\subsection{Identification of strong gravity domain in FLRW background}
%
%
%
%
\par\noindent
The Ricci scalar for the FLRW metric is given by
%
%
%
%
\beq
\mathcal{R}\,=\,\f{6}{a^2\,c^2}\,\bigg(a\ddot{a}+\,\dot{a}^{2}\bigg)
\label{ricci_flrw_1}
\eeq
where $\dot{} \equiv d/dt$. Using the Einstein's equation for FLRW metric one obtains the Friedmann equations as
%
%
%
%
%
%
\begin{equation}
 \left.\begin{aligned}
 &\f{\ddot{a}}{a}\,=\,-\,\f{4\pi\,G}{3\,c^2}\,\,(\rho+\,3\,p)\\
 & H^2=\bigg(\f{\dot{a}}{a}\bigg)^2\,=\,\f{8\pi\,G\rho}{3\,c^2}
\label{friedmann_eq_1}
       \end{aligned}
\,\,\,\,\,\,\,\, \right\}
\end{equation}
where $\o=\f{p}{\rho}$ is the parameter specifying the
equation of state, $\rho$ is  the proper energy density, $p$ the pressure in the rest frame of the proper fluid, and $H$ the Hubble parameter. Using the above equations, we write the curvature scalar as following
%
%
%
%
%
%
\beq
\mathcal{R}\,=\,\f{8\pi\,G\,\rho}{c^4}\,\,(1-3\,\o)~.
\label{ricci_energy_1}
\eeq
\par\noindent
It is well known that $\rho_{inf}(\sim\,10^{60}\,{\rm Gev}^4)\gg\,\rho_{RD}\,(\sim\,{\rm MeV}^4)\,>\,\rho_{MD}\,(\sim\,{\rm eV}^4)$, where $\rho_{inf},\,\rho_{RD},\,\rho_{MD}$ are the proper energy densities of inflationary, RD, and MD phases of the Universe, respectively \cite{Baumann:2009ds} .
Now using $\rho_{inf},\,\rho_{RD},\,\rho_{MD}$ and ($\o=\,-1,\, 1/3, 2/3$) for inflationary, RD and MD phases, respectively, one can perceive that the curvature scalar corresponding to the inflationary epoch is much larger than the curvature scalar for the RD and MD eras.
\\

%

%
%
%
\subsection{GUP modified KG equation in the FLRW background}
%
%
%
%
%
%
%
%
\par\noindent
In the background of the FLRW metric (\ref{metric_flrw}), the scalar field equation (\ref{scalar1}) takes the form
\beq
-\f{1}{c^2\,a^3(t)}\,\p_t\bigg[a^3(t)\,\p_t\,\Phi\bigg]+\f{1}{a^2(t)r^2}\p_r\bigg[r^2\,\p_r\,\Phi\bigg]+\f{1}{a^2(t)r^2\,{\rm sin}\theta}\p_{\theta}\bigg[{\rm sin}\theta\,\p_{\theta}\,\Phi\bigg]\nonumber\\
+\f{1}{a^{2}(t)r^2{\rm sin}^{2}\theta}\p_{\phi}^{2}\,\Phi\,-\,\f{m^2\,c^2}{\hbar^2}\Phi\,=\,0~.
\label{kg_flrw_1}
\eeq
\par\noindent
Furthermore, assuming $\Phi(t,r,\theta,\phi)=\t(t)\psi(r)Y_{l,m}(\theta, \phi)$ the 
separated time and space parts are
%
\beq
\hbar^2\,\f{d}{dt}\bigg[a^3(t)\,\f{d}{dt}\,\t(t)\bigg]+m^2c^4\,a^3(t)\,\t(t)+\,E^2\,a(t)\t(t)=0
\label{temporal_flrw}\\
-\f{\hbar^2}{r^2}\,\f{d}{dr}\bigg[r^2\,\f{d}{dr}\,\psi(r)\bigg]+\f{\hbar^2}{r^2}\,l(l+1)\,\psi(r)-\f{E^2}{c^2}\,\psi(r)=0~.
\label{spatial_flrw}
\eeq
As before, we focus on the radial part (\ref{spatial_flrw}) and incorporate the GUP to linear order in $\a$ to get
%
%
%
%
\beq
\f{2\hbar^2}{r}\,\f{d\psi(r)}{dr}\,+\hbar^2\,\f{d^2\psi(r)}{dr^2}+\f{4\,i\a\, \hbar^3}{r}\,\f{d^2\psi(r)}{dr^2}+2\,i\a\,\hbar^3\,\f{d^3\psi(r)}{dr^3}-\f{\hbar^2}{r^2}\,l(l+1)\,\psi+\f{E^2}{c^2}\,\psi(r)\,=\,0~.
\label{spatial_gup_flrw}
\eeq
\par\noindent
It is noteworthy that the above equation holds for all the epochs of the Universe including strong and weak gravity regimes. This implies that any space quantization obtained from the above will also hold for strong and weak gravity regimes. 
%
%
%
%
%
%
%
Now, we study Eq.(\ref{spatial_gup_flrw}) under the small and large $r$ condition in our subsequent analysis. 
\par\noindent
For small values of $r$, i.e, $r\to 0$, Eq.(\ref{spatial_gup_flrw}) reads
\beq
\,2\,i\a\,\hbar\,\f{d^2\psi(r)}{dr^2}\,
+ \f{d\psi(r)}{dr}\,
-\,\f{l(l+1)}{2r}\,\psi(r)\,=\,0~.
\label{gup_flrw_1}
\eeq
\par\noindent
For large values of $r$, i.e., $r\to \infty$, Eq. (\ref{spatial_gup_flrw}) reads
\beq
2\,i\a\,\hbar^3\,\f{d^3\psi(r)}{dr^3} 
+ \hbar^2\,\f{d^2\psi(r)}{dr^2} 
+\f{E^2}{c^2}\,\psi(r)\,=\,0~.
\label{large_gup_flrw}
\eeq
Solving Eq.(\ref{gup_flrw_1}), we get
\beq
\psi(r)=
\,C_1\frac{i \, _1F_1\left(1-\frac{l^2}{2}-\frac{l}{2}\,;\,\,2\,\,;\,\,\frac{i r}{2 \alpha  \hbar }\right)}{2 \alpha  \hbar}\,r\,
+
\,C_2\,\, G_{1,2}^{2,0}\left(-\frac{i r}{2 \alpha  \hbar }\bigg{|}
\begin{array}{c}
 \frac{1}{2} \left(l^2+l+2\right) \\
 0,1 \\
\end{array}
\right)\,
\label{sol_r_small_frw}
\eeq
where $_1F_1\left(1-\frac{l^2}{2}-\frac{l}{2}\,;\,\,2\,\,;\,\,\frac{i r}{2 \alpha  \hbar }\right)$ is a confluent hypergeometric function and $G_{1,2}^{2,0}$ denotes a Meijer-G function. We set the boundary conditions for small r to be
\be
\psi(r\,\to 0)\,=\,0\,\,\,\,\,\,\,\,\,\,\,\,\,\,\, \psi(r=L)=0~.\label{bdcs_frw}
\ee
The first condition demands that
\beq
\frac{C_2}{\Gamma \bigg[\frac{1}{2} \left(l^2+l+2\right)\bigg]}\,=\,0\,\,\implies \,\, C_2\,\,=\,\,0~.
\label{bdc_small_frw}
\eeq
Implementing the second boundary condition, we obtain
\beq
\psi(L)\,=\,\, _1F_1\left(1-\frac{l^2}{2}-\frac{l}{2}\,;\,\,2\,\,;\,\,\frac{i L}{2 \alpha  \hbar }\right)\,\,=\,\,0~.
\label{hyper_frw}
\eeq
%
%
%
%
%
Restricting ourselves to the test field $\Phi$ (and hence $\psi(r)$) 
preserving the homogeneity and isotropy of the background spacetime, i.e. for $l=0$, we have
%
{
$_1F_1\left(1-\frac{l^2}{2}-\frac{l}{2}\,;\,\,2\,\,;\,\,\frac{i L}{2 \alpha  \hbar }\right)\,=\,\frac{2 i\, \alpha\hbar} {L} \left(e^{\frac{i L}{2 \alpha\hbar}}-1 \right)$, so Eq. (\ref{hyper_frw}) becomes
\beq
&&\frac{2 i\, \alpha\hbar} {L} \left(e^{\frac{i L}{2 \alpha\hbar}}-1 \right)\,=\,0\nonumber\\
\implies\, &&\bigg[\bigg\{{\rm cos}\bigg(\f{L}{2\a\hbar}\bigg)-1\bigg\}+i\,{\rm sin}\bigg(\f{L}{2\a\hbar}\bigg)\bigg]\,=\,0
\label{hyper_l0_1}
\eeq
}
\par\noindent
{and we end up with a length quantization condition of the form 
\be
L= 4n\pi \alpha_{0}l_{pl}
\label{quantz_small_r}
\ee
with $n\,\in\,\mathbb{N}$.}
%
%
%
%
%
\par\noindent
We now turn our attention to large $r$.  
It is easily seen that Eq. (\ref{large_gup_flrw}) is identical to that for a particle in a box in flat spacetime as in Ref. \cite{Ali:2009zq}.
For a detailed analysis, we refer the reader
to Eq. (11) of Ref. \cite{Ali:2009zq} and the 
subsequent discussion. 
Solving Eq. (\ref{large_gup_flrw}) and
keeping terms up to the leading order in $\a$, we obtain a solution for $\psi(r)$ of the form
\beq
\psi(r)=\,A\,e^{ik'\,r}+\,B\,e^{-ik''\,r}+\,Ce^{\f{i\,r}{2\a\hbar}}
\label{large_r_frw_1}
\eeq
where, $k'=k(1+k\a\hbar)$, $k''=k(1-k\a\hbar)$, and $k^2=\f{E^2}{c^2\hbar^2}$. %
It should be pointed out that in the limit $\a \to 0$, Eq. (\ref{large_gup_flrw}) reduces to the one-dimensional Schr$\ddot{{\rm o}}$dinger equation.
The appearance of the first term in Eq. (\ref{large_gup_flrw}) is due to the GUP and as a consequence, an additional oscillatory solution appears in Eq. (\ref{large_r_frw_1}). For this to go away in the 
$\alpha\rightarrow 0$ limit, we require
$\lim_{\a \to 0}\,|C|\,=\,0$. 
Furthermore, for simplification we consider $A$ to be real and absorb any phase factor of $A$ within $\psi(r)$. 
\par\noindent
Next we set the boundary conditions in the form
$\psi(r=L_1)\,=\,\psi(r=L_2)\,=\,0$. We impose these boundary conditions in Eq. (\ref{large_r_frw_1}) and consider $L=\,L_2\,-\,L_1$, with  $L$ to be the characteristic length of the space where the particle resides  and is a small quantity. We also expand ${\rm exp}[i(\ldots) L)]$
and keep terms up to the leading order in $L$. 
Thus, we  obtain 
\beq
e^{\f{iL_2}{2\a\hbar}}\,(i\, L)\bigg[A\,e^{ik'L_1}\,\bigg(\f{1}{2\a\hbar}-\,k'\bigg)+B\,e^{-ik''L_1}\,\bigg(\f{1}{2\a\hbar}+\,k''\bigg)\bigg]\,=\,0~.
\label{bdc_large_1}
\eeq
From the above equation, three possibilities emerge
\begin{equation}
 \left.\begin{aligned}
 &{\rm Case\,1:} \,\,\,\,\,\,\,\,\,\,\,\,\,  &&e^{\f{iL_2}{2\a\hbar}}=0\\
 &{\rm Case\,2:}\,\,\,\,\,\,\,\,\,\,\,\,\,     &&  L =0\\
 &{\rm Case\,3:}\,\,\,\,\,\,\,\,\,\,\,\,\, &&A\,e^{ik'L_1}\,\bigg(\f{1}{2\a\hbar}-\,k'\bigg)+B\,e^{-ik''L_1}\,\bigg(\f{1}{2\a\hbar}+\,k''\bigg)=0
       \end{aligned}
 \right\}~.
 \label{cases_frw_large}
\end{equation}
As can be seen, Case 1 does not give a condition on
$L$.
%
Case 2 does not turn out to be  useful 
either because we chose the boundaries such that $L \neq 0$.
%
%
%
%
%
Finally, Case 3 yields
\beq
B\,=\,A\,e^{2i\,k\,L_1}\,\,\bigg[\,\,\f{k+k^2\,\a\hbar-\f{1}{2\a\hbar}}{k-k^2\,\a\hbar+\f{1}{2\a\hbar}}\,\,\bigg]~.
\label{b_a_1}
\eeq
\par\noindent
Now we explore Case 3 by substituting Eq. (\ref{b_a_1}) in  Eq.(\ref{large_r_frw_1}), and using $r=L_1$ and  $C=\,|C|\,e^{-i\,\theta_C}$,  we obtain
\beq
\f{2\,A\,k}{k-k^2\,\a\hbar+\f{1}{2\a\hbar}}\,=\,e^{i\bigg(\f{L_1}{2\a\hbar}-k L_1-\theta_C\bigg)}(|C|\,i\a\,\hbar\,L_1\,k^2-|C|)
\label{case3_1}
\eeq
%
%
%
%
%
%
%
where in the above equation, we have expanded $e^{i\,k^2\a\hbar L_1}$ and kept terms up to the linear order in $\a$, i.e., $e^{-ik^{2}\alpha\hbar L_{1}} \approx (1 -ik^{2}\alpha\hbar L_{1})$ since  $k^{2}\alpha\hbar L_{1}$ is small. Then, in Eq. (\ref{case3_1}), this term is multiplied with $|C|$ which is also small (as explained three (3) lines below Eq. (\ref{large_r_frw_1})) and, thus, we can neglect the first term in the RHS of Eq. (\ref{case3_1}) compared to the second one. 
%
%
%
%
\beq
|C|\,{\rm cos}\bigg[\f{L_1}{2\a\hbar}-k L_1-\theta_C\bigg]+i\,|C|\,{\rm sin}\bigg[\f{L_1}{2\a\hbar}-k L_1-\theta_C\bigg]\,=\,-\,\,\f{2\,k\,A}{k-k^2\,\a\hbar+\f{1}{2\a\hbar}}~.
\label{case3_2}
\eeq
Equating the imaginary parts from both sides of the above equation, we  obtain
\beq
\f{L_1}{2\a\hbar}-k L_1-\theta_C\,=\,n_1\,\pi
\label{quant_L1}
\eeq
where $n_1 \,\in\, \,\mathbb{N}$. 
%
%
%
\\
\par\noindent
Now, following similar analysis as above, we substitute Eq. (\ref{b_a_1}) 
in Eq. (\ref{large_r_frw_1}) and use $r=L_2$. Thus, we obtain
\beq
A\,\bigg[\,\,1+\,e^{-2i\,kL}\,\bigg\{\f{k+k^2\,\a\hbar-\f{1}{2\a\hbar}}{k-k^2\,\a\hbar+\f{1}{2\a\hbar}}\bigg\}\,\,\bigg]=\,-|C|e^{i\bigg(\f{L_2}{2\a\hbar}\,-k\,L_2-\,\theta_C\bigg)}\,\,e^{-ik^2\a\hbar\,L_2}~.
\label{case_3_add}
\eeq
It can be noted that for $\a \to 0$ the factor $\bigg\{\f{k+k^2\,\a\hbar-\f{1}{2\a\hbar}}{k-k^2\,\a\hbar+\f{1}{2\a\hbar}}\bigg\}\,\to\,-1$. Therefore, in the $\a \to 0$ limit both sides of the above equation vanish when $kL=p\pi$ and $|C|\,=0$, where $p$ $\in\, \mathbb{N}$. Thus, when $\a \neq 0$, we can expect that $k  L=p\,\pi+\delta$, where $\delta$ $\in\,\mathbb{R}$, and $\lim_{\a\to 0}\delta =0$. 
This implies that $\d$ has to be proportional to  $\a^{q}$\, where $q \in\,\mathbb{R}$ with $q\,>\,0$.
Now, following the previous analysis, Eq. (\ref{case_3_add}) further reduces to
\beq
-\f{2\,k\,A}{k-k^2\,\a\hbar+\f{1}{2\a\hbar}}\,=\,|C|\,e^{i\bigg(\f{L_2}{2\a\hbar}-k L_2-\theta_C\bigg)}\,-\,2ik\,A\, L\, \bigg[\,\,\f{k+k^2\,\a\hbar-\f{1}{2\a\hbar}}{k-k^2\,\a\hbar+\f{1}{2\a\hbar}}\,\,\bigg]~.
\label{case3_3}
\eeq
Equating the RHS of Eq. (\ref{case3_3}) with the LHS of Eq.  (\ref{case3_2}), we  obtain
\beq
|C|\,e^{i\big(\f{L_2}{2\a\hbar}-k L_2-\theta_C\big)}\,\bigg[1-e^{-i  L\big(\f{1}{2\a\hbar}-k\big)}\bigg]\,=\,2ikA\, L\,\bigg[\,\,\f{k+k^2\,\a\hbar-\f{1}{2\a\hbar}}{k-k^2\,\a\hbar+\f{1}{2\a\hbar}}\,\,\bigg]~.
\label{case3_4}
\eeq
Expanding $e^{-iL\big(\f{1}{2\a\hbar}-k\big)}$ with respect to $L$ and keeping terms up to the leading order in $L$, we obtain
\beq
|C|\,{\rm cos}\bigg[\f{L_2}{2\a\hbar}-k L_2-\theta_C\bigg]+i\,|C|\,{\rm sin}\bigg[\f{L_2}{2\a\hbar}-k L_2-\theta_C\bigg]\,=\,\,\f{2k\,A}{\big(\f{1}{2\a\hbar}-k\big)}\,\,\bigg[\,\,\f{k+k^2\,\a\hbar-\f{1}{2\a\hbar}}{k-k^2\,\a\hbar+\f{1}{2\a\hbar}}\,\,\bigg]~.
\label{case3_5}
\eeq
Equating the imaginary parts from both sides of the above equation, we obtain the following condition
\beq
\f{L_2}{2\a\hbar}-k L_2-\theta_C\,=\,n_2\,\pi
\label{quant_L2}
\eeq
where $n_2 \,\in\, \mathbb{N}$. Subtracting Eq. (\ref{quant_L1}) from Eq. (\ref{quant_L2}), we get
\beq
\f{L}{2\a\hbar}\,=\,(n_2-n_1)\pi\,+\,k\, L
\label{quant_large_1}
\eeq
%
%
%
%
%
where $n_2>\,n_1$ since $L_2>\,L_1$. Following the previous discussion, we use $k L = p\,\pi+\delta$ in the above equation and end up with a length quantization condition of the form 
\beq
&&\f{ L}{2\a\hbar}\,=\,\underbrace{(n_2-n_1+ p)}_{n}\,\pi +\d\nonumber\\
&& L\,=\,2\,n\,\a\,
\hbar \pi +2\a\hbar\, \d
\label{qunatiz_frw_large}
\eeq
where $n$ $\in\,\,\mathbb{N}$. It was mentioned earlier that $\d$ ought to be proportional to $\a^{q}$ which leads the last term of Eq. (\ref{qunatiz_frw_large}) to be of the order of a higher power in $\a$. 
Therefore, we discard this term and obtain a length  
quantization condition of the form
%
\be
L= 2n\pi \alpha_{0}l_{pl}~.
\label{qunatiz_frw_large1}
\ee
From the imaginary parts of Eqs. (\ref{case3_2}) and (\ref{case3_5}), the sine functions are determined, and this leads to ${\rm cos}\big[\f{L_1}{2\a\hbar}-k L_1-\theta_C\big]={\rm cos}\big[\f{L_2}{2\a\hbar}-k L_2-\theta_C\big]=\pm\,1$, by the well known trigonometric identity. Therefore, the cosine functions in Eqs. (\ref{case3_2}) and (\ref{case3_5}) will also give rise to similar quantization condition as in Eq. (\ref{qunatiz_frw_large}).
Furthermore, the real parts of Eqs. (\ref{case3_2}) and (\ref{case3_5}) will determine the relation between the two constants $A$ and $C$. 
{It should  be noted that the length quantization condition given in Eq. (\ref{qunatiz_frw_large1}) is consistent with the length quantization condition given in Eq. (\ref{quantz_small_r})}.
\par\noindent
The results of this section show that despite the time dependence of the background spacetime, the discreteness of space in terms of the same fundamental unit remains in all epochs of the Universe. 
This shows the robustness of our results
as well as the fundamental nature of the scale of discreteness.
%
%
%
%
%
\section{Discussion}\label{discussions}
%
%
%
%
%
%
\par\noindent
In this paper we have derived the discreteness of measured
lengths in strong gravitational fields in a number of BH and 
cosmological spacetimes. 
In particular, we worked with the Schwarzschild and RN  BH spacetimes as well as the FLRW cosmological spacetime. 
For the BH spacetime, the strong gravity region is close to the singular region of the BH whereas for the FLRW spacetime it is in the inflationary epoch of the Universe. Remarkably, the derived discreteness turns out to be independent 
of parameters such as the Schwarzschild mass, the RN charge, the KG mass or for that matter any particular epoch during the evolution of our Universe.
{In addition}, it agrees with the corresponding results derived for  
zero gravity (flat spacetime) and weak gravity (weakly curved spacetime) 
regimes and employing test fields of various spins  
\cite{Ali:2009zq,Ali:2011fa,Deb:2016psq}. 
%
%
%
{Furthermore, all length quantization conditions independently of the gravitational background are of the form 
$L \sim n\pi \alpha_{0}l_{pl}$ modulo a natural number that depends on the specific gravitational background under study.}
Finally, as in many previous works \cite{Das:2008kaa, Das:2010sj, Das:2009hs,Ali:2009zq,Das:2010zf,Deb:2016psq},
our results suggest the existence of a ``new'' length scale $l_{new} \equiv \alpha_{0}l_{Pl}$. 
A recent theoretical work suggests $\alpha_0 ={\cal O}(1)-{\cal O}(10)$ \cite{Scardigli:2016pjs}. However, since no values of $\alpha_0$ are experimentally ruled out, we leave the possibility open for $\alpha_0$ to assume higher values, resulting in an intermediate length scale between the Planck and the electroweak length scales. This would lead to length quantization conditions of the form  $L\sim n\pi l_{new}$.
%
%
%
%
%
%
%
%
This points towards the fundamental nature and in fact the 
potential universality of the discreteness of measured spaces.
As we saw this follows directly from the application of GUP, 
which is another
robust predictions of most candidate theories of QG. 
As for the concrete computation to arrive at the above results,
we studied the massive KG equation in the background of a spacetime 
with strong gravity. 
Since QG effects cannot be ignored in this regime,
we included this by implementing GUP in the KG equation. 
%
%
%
%
%
%
%
{The form of GUP that we employed in our analysis is not manifestly Lorentz invariant. Therefore}
one may ask as to why we did not try to implement a 
relativistic version of GUP, e.g. as in Refs. \cite{Hossenfelder:2006cw,Kober:2010sj,Capozziello:1999wx,Quesne:2006is,Todorinov:2018arx,Bosso:2020fos,Bosso:2020jay}. 
%
%
%
It should be stressed that while the aforesaid relativistic version of GUP is Lorentz covariant in
flat spacetimes, it still needs to be extended to curved spacetimes. 
Furthermore, the form of GUP utilized here can be viewed as an ``effective theory" from a fully covariant theory as described in Ref. \cite{Bosso:2018ufr}.
%
%
%
Of course, our results may 
undergo further modifications due to higher order QG effects 
arising from the higher order terms in $\alpha$ which we have ignored. %
{We hope to explore the implications of our results in the future.}
\par\noindent

%
%


\section{Acknowledgement}
\par\noindent
The work was supported by the Natural Sciences and Engineering Research Council of Canada. A. Das would like to thank D. Kothawala for useful discussions. The authors would like to thank the Referees for their constructive comments. This research was supported by the Quantum Major Innovation Fund Project, funded by the Government of Alberta.
%
%
%
%
%
%
%
\appendix
%
%
%
%
\section{Equation of motion for a scalar field in D-dimensional spacetime}\label{app_1}
%
%
%
%
%
\par\noindent
The Klein-Gordon equation in a curved background is
\begin{eqn}
\Box \Phi=\f{-1}{\sqrt{-g^{(D)}}}\p_{M}\bigg[\sqrt{-g^{(D)}}g^{MN}\p_{N}\Phi\bigg]\label{scalar2}
\end{eqn}
\par\noindent
with
\begin{eqn}
\sqrt{-g^{(D)}}=r^{D-2}\,({\rm sin}\chi_1)^{D-3}\, ({\rm sin}\chi_2)^{D-4}.....({\rm sin}\chi_{D-3})\label{gen_ang_1}~.
\end{eqn}
In the above equation $(M,\,N)$ runs from $(0,1,...D-1)$. We write the mode solution for the scalar field as following
\beq
\Phi(t,r,\chi_1,\chi_2,....,\chi_{D-2})=\t(t)\psi(r)\,Y_{l,m,n..}(\chi_1,\chi_2,\chi_3,...,\chi_{D-2})\label{solution1}~.
\eeq

\par\noindent
The scalar field equation given by Eq. (\ref{scalar1}) can be explicitly written as
\begin{eqn}
&\f{1}{c^2\sqrt{-g^{(D)}}}\p_{t}\bigg[\sqrt{-g^{(D)}}g^{tt}\p_{t}\Phi\bigg]+\f{1}{\sqrt{-g^{(D)}}}\p_{r}\bigg[\sqrt{-g^{(D)}}g^{rr}\p_{r}\Phi\bigg]\nonumber\\
&+\f{1}{\sqrt{-g^{(D)}}}\p_{\chi_1}\bigg[\sqrt{-g^{(D)}}g^{\chi_1\chi_1}\p_{\chi_1}\Phi\bigg]+\f{1}{\sqrt{-g^{(D)}}}\p_{\chi_2}\bigg[\sqrt{-g^{(D)}}g^{\chi_2\chi_2}\p_{\chi_2}\Phi\bigg]+\ldots-\f{m^2c^2}{\hbar^2}\Phi=0~.\label{scalar3}
\end{eqn}
Employing Eq. (\ref{gen_ang_1}) and utilizing the metric as given in Eq. (\ref{genmetric1}), we obtain 
\beq
&-\f{1}{c^2f(r)}\p_{t}^{2}\Phi+\f{1}{r^{D-2}}\p_r \bigg[r^{D-2}f(r)\p_r \Phi \bigg]+\f{1}{r^2 ({\rm sin}\chi_{1})^{D-3}}\,
\p_{\chi_1}\bigg[({\rm sin}\chi_{1})^{D-3}\,\p_{\chi_1}\Phi\bigg]\nonumber\\
&+\f{1}{r^2 ({\rm sin}^2\chi_{1})}\,\f{1}{({\rm sin}\chi_{2})^{D-4}}\, \p_{\chi_2}\bigg[({\rm sin}\chi_{2})^{D-4}\, \p_{\chi_2} \Phi \bigg]+\ldots-\f{m^2c^2}{\hbar^2}\Phi=0~.
\label{scalar4}
\eeq
Then, we can write the solution for the scalar field mode in the form
\beq
\Phi(t,r,\chi_1,\chi_2,\ldots,\chi_{D-2})=\t(t)\psi(r)\,Y_{l,m,n,\ldots}(\chi_1,\chi_2,\chi_3,\ldots,\chi_{D-2})\label{solution1}
\eeq
and, thus, Eq. (\ref{scalar4}) becomes
\beq
&-\f{1}{c^2f(r)}\p_{t}^{2}[\t(t)]\psi(r)Y_{l,m,n,\ldots}(\chi_1,\chi_2,\chi_3,\ldots,\chi_{D-2})+\f{1}{r^{D-2}}\p_r \bigg[r^{D-2}f(r)\p_r \psi(r) \bigg]\t(t)Y_{l,m,n, \ldots}(\chi_1,\chi_2,\chi_3,\ldots,\chi_{D-2})+\nonumber\\
&\f{1}{r^2 ({\rm sin}\chi_{1})^{D-3}}\,
\p_{\chi_1}\bigg[({\rm sin}\chi_{1})^{D-3}\,\p_{\chi_1}Y_{l,m,n,\ldots}(\chi_1,\chi_2,\chi_3,\ldots,\chi_{D-2})\bigg]\psi(r)\t(t)\nonumber\\
&+\f{1}{r^2 ({\rm sin}^2\chi_{1})}\,\f{1}{({\rm sin}\chi_{2})^{D-4}}\, \p_{\chi_2}\bigg[({\rm sin}\chi_{2})^{D-4}\, \p_{\chi_2} Y_{l,m,n,\ldots}(\chi_1,\chi_2,\chi_3,\ldots,\chi_{D-2}) \bigg]\psi(r)\t(t)+\nonumber\\
&\ldots-\f{m^2c^2}{\hbar^2}\, \t(t)\psi(r)Y_{l,m,n, \ldots}(\chi_1,\chi_2,\chi_3,\ldots,\chi_{D-2})=0~.
\label{scalar5}
\eeq
By dividing both sides of the above equation with $\t(t)\,\psi(r)\,Y_{l,m,n,\ldots}(\chi_1,\chi_2,\chi_3,\ldots,\chi_{D-2})$, we obtain

\beq
&-\f{1}{c^2f(r)\t(t)}\p_{t}^{2}[\t(t)]+\f{1}{r^{D-2}}\p_r \bigg[r^{D-2}f(r)\p_r \psi(r) \bigg]\f{1}{\psi(r)}+\nonumber\\
&\f{1}{r^2 ({\rm sin}\chi_{1})^{D-3}Y_{l,m,n,\ldots}}\,
\p_{\chi_1}\bigg[({\rm sin}\chi_{1})^{D-3}\,\p_{\chi_1}Y_{l,m,n,\ldots}(\chi_1,\chi_2,\chi_3,\ldots,\chi_{D-2})\bigg]\nonumber\\
&+\f{1}{r^2 ({\rm sin}^2\chi_{1})}\,\f{1}{({\rm sin}\chi_{2})^{D-4}}\f{1}{Y_{l,m,n,\ldots}}\, \p_{\chi_2}\bigg[({\rm sin}\chi_{2})^{D-4}\, \p_{\chi_2} Y_{l,m,n,\ldots}(\chi_1,\chi_2,\chi_3,\ldots,\chi_{D-2}) \bigg]+\ldots-\f{m^2c^2}{\hbar^2}=0
\label{scalar6}
\eeq
which is rewritten as
\beq
&-\f{r^2\hbar^2}{c^2f(r)\t(t)}\p_{t}^{2}[\t(t)]+\f{\hbar^2r^2}{r^{D-2}}\p_r \bigg[r^{D-2}f(r)\p_r \psi(r) \bigg]\f{1}{\psi(r)}-m^2\, r^2\,c^2=\nonumber\\
&-\hbar^2\,\bigg[\f{1}{({\rm sin}\chi_{1})^{D-3}Y_{l,m,n,\ldots}}\,
\p_{\chi_1}\bigg[({\rm sin}\chi_{1})^{D-3}\,\p_{\chi_1}Y_{l,m,n,\ldots}(\chi_1,\chi_2,\chi_3,\ldots,\chi_{D-2})\bigg]\nonumber\\
&+\f{1}{({\rm sin}^2\chi_{1})}\,\f{1}{({\rm sin}\chi_{2})^{D-4}}\f{1}{Y_{l,m,n,\ldots}}\, \p_{\chi_2}\bigg[({\rm sin}\chi_{2})^{D-4}\, \p_{\chi_2} Y_{l,m,n, \ldots}(\chi_1,\chi_2,\chi_3,\ldots,\chi_{D-2}) \bigg]+\ldots\bigg]~.
\label{scalar7}
\eeq
In Eq. (\ref{scalar7}), we equate both sides with $\hbar^2\,l(l+1)$ and, therefore, 
the RHS which is the angular part becomes 
\beq
&\f{\hbar^2}{({\rm sin}\chi_{1})^{D-3}Y_{l,m,n,\ldots}}\,
\p_{\chi_1}\bigg[({\rm sin}\chi_{1})^{D-3}\,\p_{\chi_1}Y_{l,m,n,\ldots}(\chi_1,\chi_2,\chi_3,\ldots,\chi_{D-2})\bigg]\nonumber\\
&+\f{\hbar^2}{({\rm sin}^2\chi_{1})}\,\f{1}{({\rm sin}\chi_{2})^{D-4}}\f{1}{Y_{l,m,n,\ldots}}\, \p_{\chi_2}\bigg[({\rm sin}\chi_{2})^{D-4}\, \p_{\chi_2} Y_{l,m,n,\ldots}(\chi_1,\chi_2,\chi_3,\ldots,\chi_{D-2}) \bigg]+\ldots=\,-\,\hbar^2\,l(l+1)
\label{scalar_8_app}
\eeq
\par\noindent
and, consequently, the LHS part of Eq. (\ref{scalar7}) will be
\beq
-\f{\hbar^2r^2}{c^2f(r)\t(t)}\p_{t}^{2}[\t(t)]+\f{\hbar^2r^2}{r^{D-2}}\p_r \bigg[r^{D-2}f(r)\p_r \psi(r) \bigg]\f{1}{\psi(r)}-m^2\, r^2\,c^2= \hbar^2\,l(l+1)~.
\label{scalar_9_app}
\eeq
\par\noindent
In $D=4$ spacetime dimensions, the angular momentum operator can be identified as follows
\beq
L^2=\,-\hbar^2\bigg[\f{1}{\sin \theta}\f{\p}{\p \theta}\bigg(\sin \theta \f{\p}{\p \theta}\bigg)+\f{1}{\sin^2 \theta}\f{\p^2}{\p \phi^2}\bigg]\\
L^2\, Y_{l,m}(\theta,\phi)\,=\, \hbar^2\,l(l+1)Y_{l,m}(\theta,\phi)~.
\eeq
By adopting the analysis followed before, Eq. (\ref{scalar_9_app}) for $D=4$ reads
\beq
-\bigg(\f{\hbar^2}{c^2}\bigg)\f{1}{\t(t)}\p_{t}^{2}[\t(t)]=-\f{\hbar^2 f(r)}{r^{2}}\p_r \bigg[r^{2}f(r)\p_r \psi(r) \bigg]\f{1}{\psi(r)}+m^2\,c^2\, f(r)+\frac{\hbar^2\,l(l+1)}{r^2}f(r)~.
\label{solution_new}
\eeq
Now we equate both sides of Eq. (\ref{solution_new}) with the quantity $\f{E^2}{c^2}$ and obtain 
\beq
&-\f{1}{\t(t)}\p_{t}^{2}[\t(t)]=\f{E^2}{\hbar^2}\\
&\implies \t(t) \sim Exp\,[\pm\, iEt/\hbar]\label{solution_t}
\eeq
while the radial part reduces to
\beq
-\f{\hbar^2}{r^{2}}\f{d}{dr} \bigg[r^{2}f(r)\f{d\psi(r)}{dr} \bigg]+\bigg[\frac{\hbar^2\,l(l+1)}{r^2}-\, \f{E^2}{c^2\,f(r)}+m^2c^2\bigg]\psi(r)=0
\label{solution_r_2_app}
\eeq
where
\beq
-\f{\hbar^2}{r^{2}}\f{d}{dr} \bigg[r^{2}f(r)\f{d\psi(r)}{dr}\bigg]=-\f{f(r)\,\hbar^2}{r^{2}}\f{d}{dr} \bigg[r^{2}\f{d\psi(r)}{dr}\bigg]-\hbar^2\,f'(r)\f{d\psi}{dr}&=&-f(r)\,\hbar^2 \nabla^2\, \psi(r)-\hbar^2\,f'(r)\f{d\psi}{dr}\nonumber\\
&=&f(r)\,p_{0}^{2}\psi(r)-\hbar^2\,f'(r)\f{d\psi}{dr}~.
\label{term_2}
\eeq
%
%
%
%
%
\section{Boundary condition}\label{bdc_new}
%
%
%
%
%
\par\noindent
In order to be precise, we impose the boundary conditions of the form
\be
\psi(r=\epsilon)\,=\,0\,\,\,\,\,\,\,\, \psi(r=L)=0\label{bdcs_new}
\ee
\par\noindent
where $\epsilon$ is a radial distance close to BH singularity, i.e., $r=0$. The above boundary conditions imply
\beq
\psi(r=\epsilon)=C_1\, {\rm exp}\bigg[\f{i\epsilon}{8\a\hbar}\bigg\{1-\bigg(1-\f{8ic^2\,\a\hbar\,l(1+l)}{GM}\bigg)^{1/2}\bigg\}\bigg]+C_2\, {\rm exp}\bigg[\f{i\epsilon}{8\a\hbar}\bigg\{1+\bigg(1-\f{8ic^2\,\a\hbar\,l(1+l)}{GM}\bigg)^{1/2}\bigg\}\bigg]
\label{sol_epsilon_1}
\eeq
\beq
\psi(r=L)=C_1\, {\rm exp}\bigg[\f{iL}{8\a\hbar}\bigg\{1-\bigg(1-\f{8ic^2\,\a\hbar\,l(1+l)}{GM}\bigg)^{1/2}\bigg\}\bigg]
+C_2\, {\rm exp}\bigg[\f{iL}{8\a\hbar}\bigg\{1+\bigg(1-\f{8ic^2\,\a\hbar\,l(1+l)}{GM}\bigg)^{1/2}\bigg\}\bigg]~.
\label{sol_L_1}
\eeq
\par\noindent
Solving the above equations, we obtain
\beq
&&C_2\bigg[{\rm exp}\bigg[\f{i\epsilon}{8\a\hbar}\bigg\{1+\bigg(1-\f{8ic^2\,\a\hbar\,l(1+l)}{GM}\bigg)^{1/2}\bigg\}\bigg]\,{\rm exp}\bigg[\f{iL}{8\a\hbar}\bigg\{1-\bigg(1-\f{8ic^2\,\a\hbar\,l(1+l)}{GM}\bigg)^{1/2}\bigg\}\bigg]\nonumber\\
&&-\,{\rm exp}\bigg[\f{i\epsilon}{8\a\hbar}\bigg\{1-\bigg(1-\f{8ic^2\,\a\hbar\,l(1+l)}{GM}\bigg)^{1/2}\bigg\}\bigg]\,{\rm exp}\bigg[\f{iL}{8\a\hbar}\bigg\{1+\bigg(1-\f{8ic^2\,\a\hbar\,l(1+l)}{GM}\bigg)^{1/2}\bigg\}\bigg]\bigg]\,=\,0~.
\label{sol_epsilon_L}
\eeq
\par\noindent
Then, we simplify Eq. (\ref{sol_epsilon_L}) and we get
\beq
{\rm exp}\bigg[\f{\epsilon+L}{8\a \hbar}-\f{\pi}{2}\bigg]\,{\rm sin}\bigg[\f{L-\epsilon}{8\a\hbar}\,(1-i\a \b)^{1/2}\bigg]=0~.
\label{sol_L_2}
\eeq
\par\noindent
Finally, the above equation reduces to
\beq
\f{L+\epsilon}{8\a \hbar}\,=\,(n+1)\pi,\,\,\,\,\,\,\,\,\,n \in \mathbb{N}
\label{sol_L_3}
\eeq
\beq
\f{L-\epsilon}{8\a\hbar}\,\approx \,p\pi,\,\,\,\,\,\,\,\,\,p \in \mathbb{N}~.
\label{sol_L_4}
\eeq
\par\noindent
It can be noted that for $\epsilon \to \,0$ limit the above equation will give rise to the quantization conditions as obtained in section(\ref{metric_singular}). 
\\
%
%
%
%
%
%
%
%
%
%

%
%
%
%
%

%
%
%
\end{document}